\documentstyle[german,twoside,11pt,epsf]{article}
\newcommand{\HI}{\mbox{H$\,${\sc i}}}
\newcommand{\HII}{\mbox{H$\,${\sc ii}}}
\newcommand{\CII}{C$\,${\sc ii}}
\newcommand{\tcitemsep}{\vspace*{0.15cm} \\}
\newcommand{\tcheadsep}{\vspace*{0.25cm} \\}
\newcommand{\tcsecsep}{\vspace*{0.65cm} \\}
\newcommand{\tcautind}{\hspace*{0.3cm}}
\newcommand{\secsep}{\vspace*{0.50cm}}
\newcommand{\titsep}{\vspace*{0.20cm}}
\newcommand{\tabsep}{\vspace*{-0.15cm}}
\newcommand{\plns}{\hspace*{0.10cm}}
\newenvironment{prcap}{
\hspace*{-0.70cm} \small \begin{minipage}{16.0cm}
\setlength{\parindent}{0.5cm}
\noindent
}
{
\end{minipage}
}

\begin{document}

\thispagestyle{empty}
\begin{center}
\vspace*{2.5cm}

{\LARGE \sl Bonn/Bochum-Graduiertenkolleg Workshop} \vspace*{0.85cm}

{\huge \bf The Magellanic Clouds and} \vspace*{0.5cm}

{\huge \bf Other Dwarf Galaxies} \vspace*{2.35cm}

{\large \sf Physikzentrum Bad Honnef, Germany} \vspace*{0.55cm}

{\large \sf 19th -- 22nd January 1998}
\vspace*{2.35cm}

{\Large Edited by} \vspace*{0.45cm}

{\Large {\sc Tom Richtler} \& {\sc Jochen M. Braun}}
\vspace*{0.45cm}

{\it Sternwarte der Universit\"at Bonn, Auf dem H\"ugel 71,}

{\it D--53121 Bonn, Federal Republic of Germany}
\vspace*{3.5cm}

{\tt Shaker Verlag} \vspace*{1.2cm}

{\sl October 1998}
\end{center}
\newpage

\thispagestyle{plain}
\begin{center}
% <v> Title of your talk/poster:
{\LARGE \bf  Foreword by the Spokesman of the Graduiertenkolleg}
\end{center}
\vspace*{0.85cm}

% <v> Short title and author information for heading:
\markboth{Preface I}{Foreword by the GK Spokesman}

\setlength{\parindent}{0.5cm}
\noindent
% *** Beginning of the Main Text ***
It is a great pleasure that the 25th meeting of our Graduate Working Group,
which is being supported now in its 6th year by the Deutsche
Forschungsgemeinschaft, could become a significant international event.
It is also gratifying to see this workshop as a harvest after $5\;$years
of intensive work.
The Graduate Working Group has seen a host of guest scientists who have
visited in the framework of close collaboration, or gave lectures and seminars
during the many previous meetings.
This gathering of experts on galaxies, and particularly dwarf galaxies,
represents a small milestone in the recent past of galaxy research
at the Universities of Bonn and Bochum.

Until the late 70's, the investigation of dwarf galaxies was restricted
to a few small groups.
Although the Magellanic Clouds had been studied extensively due to their
proximity, it took a while until the term dwarf galaxy became fashionable
among astrophysicists.
There are several reasons why dwarf galaxy research is one of today's fastest
growing branches of modern astrophysics.
First, the Magellanic Clouds have always been appreciated as the ideal
astrophysical laboratory in which one can study all kinds of processes without
being hampered by distance and line-of-sight problems.
Second, since it was known early on that dwarf galaxies are so little evolved,
their relevance to cosmology was recognized.
Third, by virtue of their low mass, the evolution of their ISM and stellar
populations can be studied in the absence of internal large-scale triggers
like stellar bars or density waves.
It is therefore not surprising that the past two decades have seen a rapidly
increasing rate of related research projects and publications, accompanied by
an ever increasing number of conferences dedicated to this subject, one of
which is being reported in these proceedings.

While to the non-expert it might at first glance look as if dwarf galaxy
research is a rather specialized field, a quick look at contemporary reviews
on galaxy research demonstrates that understanding low-mass galaxies is of
utter importance to also comprehend what is going on in more massive and, in
particular, more distant ones. A prime example is the understanding of the
faint blue galaxy population at redshift 0.4 to 1.0, which has experienced a
further boom with the scrutiny of distant objects in the Hubble Deep Field.
It is obvious that in order to understand stellar systems at large distances,
it is all the more important (also for the cosmologists) to understand those
just outside our door step!

Research on the Magellanic Clouds and other dwarf galaxies had been going on
at the Universities at Bonn and Bochum for some time prior to the Bonn/Bochum
Graduate Working Group.
It was therefore quite natural to combine the experience gained in two small
research groups in order to intensify this research, while at the same time
establishing an educational programme in which graduate students could benefit
for their own career, and contribute to the progress in a very proliferous
field of research.

We are grateful to the Deutsche Forschungsgemeinschaft for financing this
Graduate Working Group, thus rendering possible this intensive educational and
research programme in a very prolific and fascinating field of research.
I would like to take this opportunity to thank Christian Br\"uns for his
invaluable help in organizing this workshop. It has been (and still is) a
great pleasure to work with the students in the Graduate Working Group.
It is their enthusiasm and diligence that, besides my natural endeavour in
this field of research, ensures permanent motivation!
% *** End of the Text ***
\vspace*{1.2cm}

\hspace*{12.30cm} \begin{tabular}{l}
Uli Klein \\
Bonn, April 1998 \end{tabular}
\newpage

\thispagestyle{plain}
\begin{center}
% <v> Title of your talk/poster:
{\LARGE \bf Foreword of the Editors}
\end{center}
\vspace*{0.85cm}

% <v> Short title and author information for heading:
\markboth{Preface II}{Editors' Foreword}

\setlength{\parindent}{0.5cm}
% *** Main body:
\noindent
After having completed the present edition of the proceedings of our workshop
on ``The Magellanic Clouds and Other Dwarf Galaxies'' we wish to express our
gratitude to all contributors.
The concept of our ``Graduiertenkolleg'' is meant to provide an efficient
framework for graduate education and research and to foster not only the
relations between students and supervisors but also between students and
distinguished representatives of the community of our field of research. 
In this sense the graduate school can be seen as an attempt to realize the old
(and often forgotten) idea of unifying teaching and research at universities. 
This edition is a handy manifestation.  

We remember quite well that, when we tried to expand our scientific ambitions
centered on the Magellanic Clouds, the extension ``... and Other Dwarf
Galaxies'' sounded meaningless to many ears.
Looking at the entire collection of reviews, talks and posters addressing
such a large variety of astrophysical issues, it now seems to us that no better
headline for merging the various interests within our graduate school could
have been found.
 
We hope that it is of interest for many researchers to have a look at the
extensive key word and object lists (see the apendices ``Subject Index'' and
``List of Objects'' on page 337 and 331 respectively) to quickly find specific
information among the wealth of topics and objects, which appear in contributed
talks and posters.
We also hope that the attractive cover (designed by Michael Hilker and Sven
Kohle) strengthens the wish of a potential reader to know what is in the book. 

We wish to thank the Shaker Verlag, namely Dr. Shaker, for giving us the
permission to make the proceedings available in free electronic form on our
webserver (see also the next preface about ``Electronic Publishing'') and
all the participants (see the ``List of Participants'' on page 313) for
their help in creating a fascinating workshop, for their contributions, and
of course for further supporting the electronic version with updates and
new links.
We very much appreciate the contribution by Joachim K\"oppen, who fell sick
shortly before the workshop and could not participate. 

Considering the statistics of incoming contributions (see electronic version,
at the URL:
{\tt http://www.astro.uni-bonn.de/$\,\,\tilde{}\,\,$webgk/ws98/techinf\_a.html\#subm}),
it is noteworthy that the first contribution was a review, and so was the
last.
The distribution of time of the arrival of the reviews was nearly homogeneous,
whereas the one for the talks and especially for the posters was peaking at
our deadline.

Many thanks go to the Deutsche Forschungsgemeinschaft as the funding
institution of the Graduiertenkolleg and also for supporting this
workshop.
\vspace*{1.2cm}

\hspace*{9.55cm} \begin{tabular}{l}
Tom Richtler \& Jochen M. Braun \\
Bonn, October 1998 \end{tabular}
\newpage

\thispagestyle{plain}
\begin{center}
% <v> Title of your talk/poster:
{\LARGE \bf Electronic Publishing}
\end{center}
\vspace*{0.85cm}

% <v> Short title and author information for heading:
\markboth{J.M. Braun - Preface III}{Electronic Publishing}

\setlength{\parindent}{0.5cm}
\noindent
In the last few years sciences have improved by new electronic means, not only
to create huge data bases and providing tools for data reduction, but also in
electronic publishing (EP) and information exchange.

One of the most powerful aids apart from E-Mail and FTP (File Transfer Protocol)
is the World Wide Web (WWW or W3), which has developed from the ``toy'' status
to a very fast publication means, useful for announcing meetings, presenting
institutes and working groups, providing the latest data about equipment, and
nowadays also for publishing.
Right from the start WWW was intended as a scientific tool developed at CERN,
but soon was adopted by the whole society, at least in some countries with
approriate and cheap telecommunication infrastructure.
Thus in addition to the tools (like browsers) developed by scientists,
commerial products arose and the language of the Web, HyperText Markup
Language - HTML, came into Babylonian trouble.
To create future standards and secure the general accessibility, the
WWW Consortium (W3C) formed, which has created the current (but still
unsupported) standards HTML$\;$4.0 and CSS$\;$2.0, and is developing further
scientificaly relevant techniques, namely XML, XSL, and DOM.

To focus on astrophysics, literature search in the good old libraries was
supported by electronic search engines, litearature data bases, and
preprint servers, e.g. ADS, CDS, and XXX, and publishing in printed journals
including the well-tried referee process was extended by electronic versions,
in different quality and accessibility.
That EP is of enhanced importance today can be seen e.g. by the controversial
panel discussion at the ``AG Tagung'' in Insbruck 1997 after the talk
``Electronic publishing in its context and in a professional perspective'' by
Andr\'e Heck (1998, Reviews in Modern Astronomy 11, 337).

Let me stress a few aspects which, from my personal point of view are import
in future: \\
 $\bullet$ We need an international referee process decoupled from
           journals or publishers and revised \hspace*{0.25cm} international
           copyright laws doing justice to sciences. \\
 $\bullet$ Scientists have to stay involved in modern techniques, i.e. knowing
           the basics of \TeX, HTML \hspace*{0.25cm} etc.,
           and every institution needs scientists especially trained in
           these subjects. \\
 $\bullet$ Continent based mirror servers should be erected providing certified
           scientific information \hspace*{0.25cm} easily and freely accessible
           for the whole international community. 

As a small step in this respect we present these proceedings in two different
versions.
One is the {\it classical} book (constructed with the typesetting system
\TeX\ \& \LaTeX ), which is published and distributed by the Shaker Verlag,
the second is the {\it electronic} version (written in HTML) on the WWW server
of the Astronomical Institutes of the University of Bonn.
The latter can be visited via the following Uniform Resource Locator (URL):
\begin{center}
{\tt http://www.astro.uni-bonn.de/$\,\,\tilde{}\,\,$webgk/ws98/cover.html}
\end{center}
and should provide the full information including updates after the day
of printing (noted in the ``Erratum'', an appendix of the electronic version)
and related URLs.
As technical modifications are possible - and even probable, people using
the HTML version may also visit the appendix ``Technical Information''
(e.g. giving recommended browsers). 
 
I hope that the electronic version will be a fruitful and frequently visited 
supplement to the printed version.
\vspace*{1.2cm}

\hspace*{11.80cm} \begin{tabular}{l}
Jochen M. Braun \\
Bonn, October 1998 \end{tabular}
\newpage

\thispagestyle{plain}
\begin{center}
{\LARGE \bf Table of Contents}
\\[0.9cm] \end{center}
\markboth{Preface III}{Table of Contents}
\begin{tabular}{l@{\quad}r}
\hspace*{14.5cm} & \\

{\large \bf Preface} \quad {\large \bf \dotfill} \qquad &
  {\large \bf iii} \tcheadsep

Foreword by the Spokesman of the Graduiertenkolleg
\quad \dotfill \quad & iii \tcitemsep

Foreword of the Editors \quad \dotfill \quad & v \tcitemsep

Electronic Publishing \quad \dotfill \quad & vii \tcitemsep

Table of Contents \quad \dotfill \quad & ix \tcsecsep

{\large \bf Review Talks} \quad {\large \bf \dotfill} \quad &
 {\large \bf 1} \tcheadsep

The Violent Interstellar Medium in Dwarf Galaxies: Atomic Gas & \\
\tcautind Elias Brinks and Fabian Walter \quad \dotfill \quad &
1 \tcitemsep

Hot Gas in the Large Magellanic Cloud & \\
\tcautind You-Hua Chu \quad \dotfill \quad &
11 \tcitemsep

Astrophysics of Dwarf Galaxies: Structures and Stellar Populations & \\
\tcautind John S. Gallagher \quad \dotfill \quad &
25 \tcitemsep

Star-Forming Regions and Ionized Gas in Irregular Galaxies & \\
\tcautind Deidre A. Hunter \quad \dotfill \quad & 37 \tcitemsep

The Law of Star Formation in Disk Galaxies & \\
\tcautind Joachim K\"oppen \quad \dotfill \quad & 45 \tcitemsep

Strange Dark Matters in Nearby Dwarf Galaxies & \\
\tcautind Mario Mateo \quad \dotfill \quad & 53 \tcitemsep

Holes and Shells in Galaxies: Observations versus Theoretical Concepts & \\
\tcautind Jan Palou\v{s} \quad \dotfill \quad & 67 \tcitemsep

Detailed Recent Star Formation Histories of Dwarf Irregular Galaxies and & \\
Their Many Uses & \\
\tcautind Evan D. Skillman, Robbie C. Dohm-Palmer, and Henry A. Kobulnicky
\quad \dotfill \quad & 77 \tcitemsep

Nearby Young Dwarf Galaxies & \\
\tcautind Trinh X. Thuan and Yuri I. Izotov
\quad \dotfill \quad & 91 \tcsecsep

{\large \bf Talks} \quad {\large \bf \dotfill} \qquad &
  {\large \bf 107} \tcheadsep

The Star Formation History of Nearby Dwarf Galaxies:
problems, methods and results & \\
\tcautind Antonio Aparicio
\quad \dotfill \quad & 107 \tcitemsep

Diffuse Hot Gas in Dwarf Galaxies & \\
\tcautind Dominik J. Bomans
\quad \dotfill \quad & 111 \tcitemsep

The Young Large-Scale Features in the Large Magellanic Cloud & \\
\tcautind Jochen M. Braun \quad \dotfill \quad & 115 \tcitemsep

A detailed view of the Magellanic Clouds in the FIR & \\
\tcautind M. Braun, R. Assendorp, Tj.R. Bontekoe, D.J.M. Kester,
          and G. Richter \quad \dotfill \quad & 121 \tcitemsep

Bow Shock Induced Star Formation in the LMC: a Large Scale View& \\
\tcautind Klaas S. de Boer \quad \dotfill \quad & 125 \tcitemsep

\end{tabular}
% ************************ Changable part!! ****************************
\begin{tabular}{l@{\quad}r}
\hspace*{14.5cm} & \\

The binary star cluster SL 538/ NGC 2006 (SL 537) & \\
\tcautind Andrea Dieball and Eva K. Grebel
\quad \dotfill \quad & 129 \tcitemsep

Tidal Dwarf Galaxies & \\
\tcautind P.-A. Duc, U. Fritze-v. Alvensleben, and P. Weilbacher
\quad \dotfill \quad & 133 \tcitemsep

The ATCA Radio-continuum Investigation of the Magellanic Clouds & \\
\tcautind Miroslav D. Filipovi\'c and Lister Staveley-Smith
\quad \dotfill \quad & 137 \tcitemsep

Tidal Dwarf Galaxies: Their Present State and Future Evolution & \\
\tcautind Uta Fritze - v. Alvensleben and Pierre-Alain Duc
\quad \dotfill \quad & 141 \tcitemsep

The star formation history of Leo$\;$I & \\
\tcautind Carme Gallart, Antonio Aparicio, Wendy Freedman, Giampaolo Bertelli,
& \\
\tcautind and Cesare Chiosi
\quad \dotfill \quad & 147 \tcitemsep

The Recent Star Formation History of the Large Magellanic Cloud & \\
\tcautind Eva K. Grebel and Wolfgang Brandner
\quad \dotfill \quad & 151 \tcitemsep

Cold dust in NGC$\;$205 & \\
\tcautind Martin Haas \quad \dotfill \quad & 155 \tcitemsep

Is the metallicity and the luminosity related for dwarf irregular galaxies? & \\
\tcautind A.M. Hidalgo-G\'amez and K. Olofsson
\quad \dotfill \quad & 157 \tcitemsep

The stellar census of the nearby BCDG VII ZW 403 with HST & \\
\tcautind Ulrich Hopp, Regina E. Schulte-Ladbeck, and Mary M. Crone
\quad \dotfill \quad & 161 \tcitemsep

\HI\ Properties of new nearby Dwarf Galaxies from the Karachentsev Catalog & \\
\tcautind W.K. Huchtmeier \quad \dotfill \quad & 165 \tcitemsep

An \HI\ Aperture Mosaic Survey of the Large Magellanic Cloud & \\
\tcautind Sungeun Kim, Lister Staveley-Smith, and Michael A. Dopita
\quad \dotfill \quad & 169 \tcitemsep

dSph Satellite Galaxies without Dark Matter: a Study of Parameter Space & \\
\tcautind Pavel Kroupa \quad \dotfill \quad & 173 \tcitemsep

Starbursts in dwarf galaxies: Blown out or blown away? & \\
\tcautind Mordecai-Mark Mac Low and Andrea Ferrara
\quad \dotfill \quad & 177 \tcitemsep

Recent Star Formation in the Wolf-Rayet BCDG MRK~1094 & \\
\tcautind David I. M\'endez, L.M. Cair\'os, C. Esteban, and J.M. V\'{\i}lchez
\quad \dotfill \quad & 181 \tcitemsep

Interpreting the \HII\ Region Luminosity Function & \\
\tcautind M.S. Oey and C.J. Clarke
\quad \dotfill \quad & 185 \tcitemsep

The first detection of H$_2$ in absorption in the LMC & \\
\tcautind Philipp Richter, Klaas S. de Boer, Dominik J. Bomans,
          Andreas Heithausen,\\
\tcautind and Jan Koornneef \quad \dotfill \quad &
          189 \tcitemsep

Chemodynamical Evolution of Dwarf Galaxies & \\
\tcautind Andreas Rieschick and Gerhard Hensler
\quad \dotfill \quad & 193 \tcitemsep

HST Study of the Stellar Populations around SN$\;$1987A & \\
\tcautind Martino Romaniello, Nino Panagia, Salvo Scuderi, and
          the SINS$\;$collaboration \quad \dotfill \quad &
          197 \tcitemsep

The Role of Galaxy Interactions in \HII\ Galaxies & \\
\tcautind Christopher L. Taylor, Elias Brinks, and Evan D. Skillman
\quad \dotfill \quad & 201 \tcitemsep

CO Emission in Low Luminosity Star Forming Galaxies & \\
\tcautind Christopher L. Taylor, Henry A. Kobulnicky, and Evan D. Skillman
\quad \dotfill \quad & 205 \tcitemsep

\end{tabular}
% ************************ Changable part!! ****************************
\begin{tabular}{l@{\quad}r}
\hspace*{14.5cm} & \\

The Outer Halo of NGC$\;$4449 & \\
\tcautind Christian Theis and Sven Kohle
\quad \dotfill \quad & 209 \tcitemsep

WFPC2 Observations of Leo$\;$A: A Young Galaxy? & \\
\tcautind Eline Tolstoy, J.S. Gallagher, A.A. Cole, J.G. Hoessel,
          A. Saha, R.~Dohm-Palmer, & \\
\tcautind E. Skillman, and M. Mateo
\quad \dotfill \quad & 213 \tcitemsep

The recent star formation history in NGC$\;$1569 & \\
\tcautind M. Tosi, M. Clampin, G. De Marchi, L. Greggio, C. Leitherer,
          and A. Nota \quad \dotfill \quad &
          217 \tcitemsep

The Violent Interstellar Medium of IC$\;$2574 & \\
\tcautind Fabian Walter, Elias Brinks, Neb Duric, J\"urgen Kerp, and Uli Klein
\quad \dotfill \quad & 221 \tcsecsep

{\large \bf Posters} \quad {\large \bf \dotfill} \quad &
 {\large \bf 225} \tcheadsep

First results of a CO survey of dwarfs & \\
\tcautind Marcus Albrecht, Roland Lemke, and Rolf Chini
\quad \dotfill \quad & 225 \tcitemsep

Preliminary Results on the Resolved Stellar Population of I$\;$Zw$\;$18 & \\
\tcautind A. Aloisi, L. Greggio, M. Tosi, M. Clampin, A. Nota, and
M. Sirianni \quad \dotfill \quad & 227 \tcitemsep

Hot subdwarfs, their kinematics and their galactic distribution & \\
\tcautind Martin Altmann, Yolanda Aguilar-S\'anchez, Klaas S. de Boer, \\
\tcautind Michael Geffert, Michael Odenkirchen, and Jacques Colin
\quad \dotfill \quad & 229 \tcitemsep

Structure and stellar content of dwarf galaxies: Surface photometry of dwarf &
\\
galaxies in the M81 \& M101 groups & \\
\tcautind Torbj{\o}rn Bremnes, Bruno Binggeli, and Philippe Prugniel
\quad \dotfill \quad & 231 \tcitemsep

\HI-Velocity-Bridges in the Magellanic Stream & \\
\tcautind Christian Br\"uns, Peter Kalberla, and Ulrich Mebold
\quad \dotfill \quad & 233 \tcitemsep

Multiband Analysis of the Brightness Distribution of a Sample of Blue & \\
Compact Dwarf Galaxies & \\
\tcautind L.M. Cair\'os and J.M. V\'{\i}lchez
\quad \dotfill \quad & 235 \tcitemsep

Recent radio and polarization observations of dwarf irregulars & \\
\tcautind Krzysztof T. Chy\.{z}y, Sven Kohle, Rainer Beck, Uli Klein, and
          Marek Urbanik \quad \dotfill \quad & 237 \tcitemsep

DeNIS Observations on the Magellanic Clouds & \\
\tcautind Maria Rosa Cioni, Cecile Loup, Harm J. Habing, and Erik R. Deul
\quad \dotfill \quad & 239 \tcitemsep

New \HI\ shells in the Milky Way & \\
\tcautind So\v{n}a Ehlerov\'a and Jan Palou\v{s} \quad \dotfill \quad &
241 \tcitemsep

The Formation and Evolution of Rich Star Clusters in the LMC: NGC 1818 & \\
\tcautind Rebecca A.W. Elson, Gerard Gilmore, Sverre Aarseth, Melvyn Davies, \\
\tcautind Steinn Sigurdsson, Basilio Santiago, and Jarrod Hurley
\quad \dotfill \quad & 243 \tcitemsep

The Local Group at High Redshift & \\
\tcautind R. Elson, R. Abraham, T. Kodama, B. Poggianti, and M.S. Oey
\quad \dotfill \quad & 245 \tcitemsep

Dwarf Galaxies in the Centaurus A Group: Interacting with their Environment?
& \\ \tcautind T. Fritz, U. Mebold, U. Klein, and S. C\^ot\'e
\quad \dotfill \quad & 247 \tcitemsep

Observational discovery of the AGB-bump in the color-magnitude diagram & \\
of nearby galaxies & \\
\tcautind Carme Gallart and Giampaolo Bertelli
\quad \dotfill \quad & 249 \tcitemsep

\end{tabular}
% ************************ Changable part!! ****************************
\begin{tabular}{l@{\quad}r}
\hspace*{14.5cm} & \\

Be Stars and the IMF of Young Clusters & \\
\tcautind Eva K. Grebel, Wolfgang Brandner, and Andrea Dieball
\quad \dotfill \quad & 251 \tcitemsep

First Results from the Magellanic Cloud Photometric Survey & \\
\tcautind Eva K. Grebel, Dennis Zaritsky, Jason Harris, and Ian Thompson
\quad \dotfill \quad & 253 \tcitemsep

The Center of the Fornax Cluster: Dwarf Galaxies, cD Halo and Globular Clusters
 & \\
\tcautind Michael Hilker, Leopoldo Infante, Markus Kissler--Patig, and
Tom Richtler \quad \dotfill \quad & 255 \tcitemsep

Identification of Be stars in the Magellanic Clouds & \\
\tcautind Dirk Hoffmann
\quad \dotfill \quad & 257 \tcitemsep

Neutral Hydrogen in a sample of selected Blue Compact Dwarf Galaxies & \\
\tcautind W.K. Huchtmeier, Gopal Krishna, and A. Petrosian
\quad \dotfill \quad & 259 \tcitemsep

NIR-Imaging Of Dwarf Galaxies & \\
\tcautind Marcus J\"utte and Rolf Chini
\quad \dotfill \quad & 261 \tcitemsep

Comparison of \HI\ and magnetic field structures in our Galaxy & \\
\tcautind H.-R. Kl\"ockner, P.M.W. Kalberla, and U. Klein
\quad \dotfill \quad & 263 \tcitemsep

A CO map of NGC$\;$4449 & \\
\tcautind S. Kohle, U. Klein, C. Henkel, and D.A. Hunter
\quad \dotfill \quad & 265 \tcitemsep

A Relation between Box/Peanut$\;$Bulges and their Satellite Systems & \\
\tcautind Rainer L\"utticke and Ralf-J\"urgen Dettmar
\quad \dotfill \quad & 267 \tcitemsep

ROSAT observations of the giant \HII\ complex N$\,$11 in the LMC & \\
\tcautind Mordecai-Mark Mac Low, Thomas H. Chang, You-Hua Chu,
          Sean D. Points, & \\
\tcautind R. Chris Smith, and Bart P. Wakker
\quad \dotfill \quad & 269 \tcitemsep

The young population of the Local Group dwarf galaxy Phoenix & \\
\tcautind D. Mart\'\i nez-Delgado and A. Aparicio
\quad \dotfill \quad & 271 \tcitemsep

The cool atomic gas in the Large Magellanic Cloud & \\
\tcautind M. Marx-Zimmer, U. Herbstmeier, F. Zimmer, J.M. Dickey,
          Staveley-Smith, & \\
\tcautind and U. Mebold
\quad \dotfill \quad & 273 \tcitemsep

CO emission toward \HI\ absorption sources in the Large Magellanic Cloud & \\
\tcautind M. Marx-Zimmer, F. Zimmer, U. Herbstmeier, Y.-N. Chin,
          J.M. Dickey, and & \\
\tcautind U. Mebold
\quad \dotfill \quad & 275 \tcitemsep

ISO-[\CII]-investigation of cool \HI\ clouds in the Large Magellanic Cloud & \\
\tcautind M. Marx-Zimmer, U. Herbstmeier, F. Zimmer, J.M. Dickey, and U. Mebold
\quad \dotfill \quad & 277 \tcitemsep

Multi-Color Surface Photometry of Blue Compact Dwarf Galaxies & \\
\tcautind K.G. Noeske, P. Papaderos, K.J. Fricke, and T.X. Thuan
\quad \dotfill \quad & 279 \tcitemsep

The Near-IR Tully-Fisher Relation for Giant and Dwarf Late-type Galaxies & \\
\tcautind Daniele Pierini and Richard Tuffs \quad \dotfill \quad &
281 \tcitemsep

The Kinematic Structure of the Supergiant Shell LMC\,2 & \\
\tcautind Sean D. Points \quad \dotfill \quad & 283 \tcitemsep

Mass Segregation in Young LMC Clusters: NGC$\;$2004 and NGC$\;$2031 & \\
\tcautind T. Richtler, P. Fischer, M. Mateo, C. Pryor, and S. Murray
\quad \dotfill \quad & 285 \tcitemsep

A photometric method to study the Wolf-Rayet content of compact \HII\ regions &
\\
in nearby galaxies & \\
\tcautind P. Royer, J.-M. Vreux, and J. Manfroid
\quad \dotfill \quad & 287 \tcitemsep

\end{tabular}
% ************************ Changable part!! ****************************
\begin{tabular}{l@{\quad}r}
\hspace*{14.5cm} & \\

Does the low-mass galaxy IC 2574 obey MOND? & \\
\tcautind F.J. S\'anchez-Salcedo and A.M. Hidalgo-G\'amez
\quad \dotfill \quad & 289 \tcitemsep

$UBVRI$ Photometry of Supergiants in Multiple Star Systems and the OB & \\
Association LH 10 & \\
\tcautind Th. Schmidt-Kaler, J. Gochermann, J. Middendorf, and W.F. Wargau
\quad \dotfill \quad & 291 \tcitemsep

Red Supergiants as a Tool for the Study of Galaxies & \\
\tcautind Th. Schmidt-Kaler \quad \dotfill \quad & 293 \tcitemsep

Photoelectric $UBV$ Photometry of Galactic Foreground and LMC Member Stars: & \\
A New Database & \\
\tcautind Th. Schmidt-Kaler, J. Gochermann, M.O. Oestreicher,
          H.-G. Grothues, & \\
\tcautind C. Tappert, A. Zaum, Th. Bergh\"ofer, and H.R. Brugger
\quad \dotfill \quad & 295 \tcitemsep

Interacting and merging processes between spirals and dwarf galaxies & \\
\tcautind Uwe Schwarzkopf and Ralf-J\"urgen Dettmar
\quad \dotfill \quad & 297 \tcitemsep

Holes and Shells in IC$\;$2574 & \\
\tcautind Fabian Walter, Elias Brinks, and Uli Klein
\quad \dotfill \quad & 299 \tcitemsep

\HI\ study of the dwarf galaxy DDO$\;$47 & \\
\tcautind Fabian Walter, Elias Brinks, and Uli Klein
\quad \dotfill \quad & 301 \tcitemsep

Bubbles around massive stars in LMC \HII\ regions & \\
\tcautind Kerstin Weis and Wolfgang J. Duschl
\quad \dotfill \quad & 303 \tcitemsep

Spatial distribution of halo high velocity clouds towards the LMC & \\
\tcautind Bernhard Wierig and Klaas S. de Boer
\quad \dotfill \quad & 305 \tcsecsep

{\large \bf Appendices} \quad {\large \bf \dotfill} \quad &
 {\large \bf 307} \tcheadsep

Conference Summary \quad \dotfill \quad & 307
\tcitemsep

Photograph of Participants \quad \dotfill \quad & 311
\tcitemsep

List of Participants \quad \dotfill \quad & 313 \tcitemsep

Final Programme \quad \dotfill \quad & 319 \tcitemsep

Glossary \quad \dotfill \quad & 323 \tcitemsep

Author Index \quad \dotfill \quad & 327 \tcitemsep

List of Objects \quad \dotfill \quad & 331 \tcitemsep

Subject Index \quad \dotfill \quad & 337 \tcitemsep

Sample Name Label \quad \dotfill \quad & EV \tcitemsep

Technical Information \quad \dotfill \quad & EV \tcitemsep

Erratum \quad \dotfill \quad & EV \tcitemsep
\end{tabular} \\[0.7cm]
\begin{center}
{\small EV -- only part of the electronic version available at the WWW URL: \\
{\tt http://www.astro.uni-bonn.de/$\,\,\tilde{}\,\,$webgk/ws98/cover.html} \\
page numbers refer to the printed version published by \\
Shaker Verlag, Aachen, ISBN 3-8265-4457-9}
\end{center}
\newpage

\thispagestyle{plain}
\begin{center}
% <v> Title of your talk/poster:
{\LARGE \bf Conference Summary}
\vspace*{0.45cm}

% <v> Short title and author information for heading:
\markboth{E.D. Skillman - Appendix A}{Conference Summary}

% <v> List of authors:
{\large \bf Evan D. Skillman}
\vspace*{0.45cm}

% <v> List of institutes:
{University of Minnesota}
\vspace*{0.85cm}

\setlength{\parindent}{0.5cm}
% *** Main body (please keep the maximum page sizes noted above):
\hspace*{-0.6cm} \begin{minipage}{14.0cm}
This is not formally a conference summary.
Foremost, the hosts are thanked and congratulated.
I then share a few impressions and a gedanken experiment is proposed.
% *** End of Abstract
\end{minipage}
\end{center} \secsep

% *** First subtitle:
{\noindent \large \bf 1. Mandatory Laudatory Comments} \titsep

\noindent
It has been my great pleasure to be a participant in this workshop.
The Bad Honnef conference center is certainly an exceptional facility
which is conducive to interaction.
The conference organizers have done a superb job in arrangements so that we
haven't had to worry about anything (with the possible exception of
the participants who still haven't made it back from K\"oln yet),
and the result has been a wealth of exchange of ideas.
I especially appreciated the poster review sessions, which helped to act as
an introduction to the many exciting results presented as posters.

This meeting represents a solid endorsement of the Graduiertenkolleg (GK),
in both its choice of science themes and its organization.
When I was a first year graduate student, Bart Bok visited the University
of Washington, and repeated over and over, ``If you want to study star
formation, study the Magellanic Clouds''.
At the time he was bemoaning the lack of a millimeter observing facility
in the southern hemisphere - something which has been rectified.
The expansion of the theme from the Magellanic Clouds to dwarf galaxies
in general has allowed the work at Bonn and Bochum to reach an even larger
audience.

Having attended an earlier meeting of the GK, I should state that my
impressions are all very positive.
The benefits of forming a concentration of studies shared between two
university departments are obvious and impressive.
The students have access to numerous first class facilities, faculty, and
many other resources.
In fact, one can say that the students have {\it enhanced} access to numerous
facilities since the resources available through the GK help to make them more
competitive for observing time.

When I attended the GK meeting in the Fall of 1995, I was impressed with the
many ideas for new projects that students were then forming.
Many of those projects have been completed and were presented at this meeting.
Good science is being done.
The GK experience has become an excellent platform from which to launch
a career in astrophysics.
\secsep

{\noindent \large \bf 2. Some Impressions} \titsep

\noindent
Some of the most successful conference summaries that I have witnessed were,
in fact, not summaries at all.
Thus, I will not attempt to summarize the whole conference, but rather to
present some of my impressions gained during the workshop.
Of course, the overall impression is that by concentrating on the nearest
galaxies one can gain insights unafforded in any other context.
Sometimes we may feel that this is lost in the rush to high redshift, but it
certainly is not lost on the participants of this workshop.
Again and again we were treated to stunning views of the Magellanic Clouds
in wavelength after wavelength.
There is a lot still to be learned from the Magellanic Clouds and even more
to be learned by pushing for comparable studies of other nearby galaxies. 
\newpage

Simply because of my recent research activities, many of my impressions are
grouped under the category of stellar populations.
First among them is that the HST is an instrument ideally suited for stellar
population studies.
The leaps forward afforded by the vast improvement over ground based resolution
has revolutionized our view of galaxies.

Nowhere is this more true than for the dwarf spheroidal galaxies.
In the last decade, our views of these ``simple'' systems has changed
enormously.
The variety of star formation histories is still begging for a simple
explanation.
The kinematic studies of literally hundreds of stars (many now with multiple
epochs) in most of the dSphs have continued to bear out Aaronson's bold claim
that these are dark matter dominated systems.
As spectra of individual stars provide detailed abundance analyses, we may
still be in for more surprises.

I was very impressed by the degree of sophistication in treatment of
color-magnitude-diagrams (CMDs) and the general agreement between independent
groups.
For example, the presentations by Aparicio and Gallart showed how the different
features in the CMDs could be used in concert to place strong constraints on
star formation histories reaching back to the earliests epochs.
Tolstoy and Hopp independently studied HST data on VII$\;$Zw$\;$403 and
(1) came to very similar conclusions about basic properties and
(2) agreed on the features in the CMD which were most difficult to fit with
    a range of reasonable models.

There was a great deal of attention paid to the detailed structure in the ISM
and its origins.
I will never again confuse the terms ``bubbles'', ``super-bubbles'', and
``supergiant shells''.

In the detailed \HI\ imaging of the Magellanic Clouds, we are confronted with
the problem of how best to deal with the high degree of complexity.
Does it make sense to divide everything into categories of ``holes'' and
``concentrations''?
Perhaps it is pertinent to reflect back on the work that Hodge conducted for
many decades, providing \HII\ region luminosity functions (which also required
making decisions about whether to divide features into components or combine
features into single entities).
We have seen that Oey has taken these \HII\ region luminosity functions and
used them to distinguish between ``saturated'' and ``unsaturated'' IMFs.
While the analogy is not sound on a physical basis, I wonder that we may not
see a similar understanding of the \HI\ holes and concentrations in the future.

Concerning ISM structure, it is evident that x-ray observations are rapidly
playing an increasing role in our understanding of the phase structure of
the ISM.
The x-ray images available today remind me of the \HI\ images that were
available a few decades ago.
As the spatial and spectral resolution and sensitivity increase, important
new insights are bound to come from this waveband.

Also, concerning the bubbles, I was disappointed that after presenting a
detailed star formation history of the Sextans$\;$A dwarf, no one asked
about the connection between the stellar population and the \HI\ hole.
I tried to check on this in preparing for this summary talk, but the limited
field of view of the HST WFPC2 observations didn't allow for a clear answer.
Recently, van Dyk has shown evidence from ground based data for a radial
gradient in stellar ages that supports a wind blown bubble model for the
central \HI\ deficit.

Moving to galaxy scales, the numerical calculations presented by Mac Low were
very impressive.
There has been endless speculation concerning the effects of star formation on
the ISM of dwarf galaxies, and, in particular, the ability of a galaxy to blow
away all of its ISM.
These new calculations appear to present a more realistic impression of what
is possible and what is not.

This leads to the connected question of what is the \HI\ that is frequently
found in the vicinity of dwarf galaxies.
I refer to this \HI\, which is usually at similar velocity but disconnected
from the normal velocity fields, as ``floppy disks''.
Two alternative origins immediately come to mind.
It could be that this material is tidal debris, left over from an interaction
or a merger.
Alternatively, this could be material which is primordial in nature and has not
yet been incorporated into the galaxy.
Within the limitations of available observing facilities, it is difficult
to design an observational test which distinguishes these two possibilities.
It is, however, a very important question.
\newpage

\begin{prcap}{\bf Table 1.}
A Summary of Impressions
\end{prcap} \tabsep

\noindent
{\small \begin{tabular}{rl}
\hline
\hline
 1) & DH TR DE ES \\
 2) & MM P BO BIP \\
 3) & ET UH FSP \\
 4) & NO AA SA \\
 5) & XR HY \\
 6) & HST BF SPS \\
 7) & IBVC BSS \\
 8) & DS EBKS \\
 9) & A H\&C LHR \\
10) & NN RFP \\
11) & FD TR PG \\
\hline
\hline
\end{tabular}}
\secsep

Finally, I have to admit that one of the most remarkable things that I heard
during the entire workshop was the story of Hunter's graduate school foreign
language exam.
To pass the exam, she was required to translate a scientific article written
in German.
In fact, she translated the article into Spanish.
I was impressed not only with the demonstration of language skills, but the
display of stubborn independence found in astronomers everywhere.

This presentation of my impressions was accompanied by a visual aid.
Uli Klein asked me to reproduce it in my conference summary, so it appears
here as Table 1.
\secsep

{\noindent \large \bf 3. A Gedanken Experiment}\titsep

\noindent
Finally, knowing that most of the workshop participants would be spending
one more night at Bad Honnef, I proposed an exercise to be carried out
after dinner.
At the time I called it an exercise, but since this is Germany, it is probably
better referred to as a gedanken experiment.

The conference participants were treated to a trip to K\"oln, the highlight
of which was a tour of the Dom.
The Dom had recently experienced some refurbishing work, as evidenced by the
many ton scaffold that was still hanging in the top of the church.

The following exercise occurred to me.
What if that scaffold had fallen and crushed some of the workshop participants?

For this exercise, I assume the following cosmology:
(1) there is a heaven,
(2) there is a single deity,
(3) when good people go to heaven the deity answers their questions, and
(4) all astronomers are good.
The purpose of the exercise is not to quibble with my assumed cosmology, but,
rather, to imagine the questions of the deceased workshop participants as they
enter heaven.

To kick off the proposed exercise, I tackled the question what if all of the
invited review speakers were done in?

I started with Jay Gallagher.
(The main reason for this was because I wasn't sure how this would go over
and Jay had already left for Munich.)
Jay's talk highlighted the degree of the interconnections between different
studies and how observations of dwarf galaxies had impact on literally every
major question in astronomy today.
Actually, I think that Jay's question was very easy.
He would ask ``What are the answers to everyone else's questions?'' 

Mario Mateo's talk emphasized the sophistication of the measurements of
the dark matter associated with dSphs and their complicated star formation
histories.
I imagine that he would be able to negotiate a second question and the two
together would be ``What is dark matter, really?'' and ``Concerning dwarf
spheroidals, where did the gas go?''

Elias Brinks has always been fascinated with detailed images of the
distribution of neutral hydrogen in galaxies.
On one hand, the energetics of the holes can be explained by the presence
of stellar clusters.
On the other hand, few of the known holes have identifiable stellar populations
in their interiors.
Elias would ask ``How are these holes formed?''
\newpage

Deidre Hunter presented detailed observations of star formation on different
scales.
In one case, she pointed to a sequence of two generations of star formation
and proposed that a third might be ready to occur.
A theory of star formation should have predictive power, and thus, Deidre
might ask ``Is there going to be a third generation of star formation?''

You-Hua Chu presented detailed comparisons of the x-ray distributions and
the gas kinematics in the Magellanic Clouds.
She questioned the assumptions of pressure equilibrium.
You-Hua would simply demand a complete picture of the multi-phase ISM in
the Large Magellanic Cloud.

Jan Palous showed us, through numerical simulations, how ISM bubbles evolve.
The calculations are simplified by taking advantage of certain symmetries.
He would like to know if 2-dimensional simulations give the correct insight
into a three dimensional world.

Trinh Thuan presented observations of SBS$\;0335-052$, which he proposes
is a young galaxy which is just forming now.
He would like to know if it is possible for galaxies to be forming in
the current epoch.

In summary, I would like to thank the organizers one last time for a truly
enjoyable workshop.
I congratulate them on their 25th GK meeting and I look forward to the 50th!
% *** End of main text
\newpage

\thispagestyle{plain}
\begin{center}
% <v> Title of your talk/poster:
{\LARGE \bf List of Participants}
\end{center}
\vspace*{0.85cm}

% <v> Short title and author information for heading:
\markboth{Appendix C}{Participants}

\setlength{\parindent}{0.5cm}
% *** Main body:
\noindent
% <v> Beginning of the first chapter of the main text:
\begin{minipage}[t]{7.6cm}

{\bf Marcus Albrecht} \plns (13)\\
% Poster: The efficiency of star formation in dwarf galaxies\\
Astronomisches Institut der\\
Ruhr-Universit{\"a}t Bochum\\
{\tt albrecht@astro.ruhr-uni-bochum.de} \titsep

{\bf Alessandra Aloisi} \plns (67)\\
% Poster: Preliminary results on the Resolved Stellar Population of the Blue
%         Compact Dwarf Galaxy Izw18\\
Dipartimento di Astronomia, \\
Universit{\'a} di Bologna\\
{\tt aloisi@astbo3.bo.astro.it} \titsep

{\bf Martin Altmann} \plns (11)\\
% Poster: Hot subdwarfs, their orbits and their galactic distribution  \\
Sternwarte der\\ Universit{\"a}t Bonn\\
{\tt maltmann@astro.uni-bonn.de} \titsep

{\bf Ralph-P. Andersen} \plns (5)\\
%Talk: The evolution of galaxies with very low mass\\
%Poster:The self-regulated evolution of dwarf galaxies\\
Max-Planck-Institut f{\"u}r Astronomie\\ (MPIA), Heidelberg\\
{\tt andersen@mpia-hd.mpg.de} \titsep

{\bf Antonio Aparicio} \plns (40)\\
%Talk: Determination of the Star Formation History in Nearby Dwarf Galaxies:
% problems, methods and results\\
Instituto de Astrof\'{\i}sica\\ de Canarias (IAC)\\
{\tt aaj@ociw.edu} \titsep

{\bf Dominik J. Bomans} \plns (12)\\
%Talk: Diffuse Hot Gas in and Evolution of Dwarf Irregular Galaxies\\
% formerly:  Astronomy Department,\\ University of Illinois\\
% formerly:  {\tt bomans@astro.uiuc.edu} \titsep
Astronomisches Institut der\\
Ruhr-Universit{\"a}t Bochum\\
{\tt bomans@astro.ruhr-uni-bochum.de} \titsep

{\bf Jochen M. Braun} \plns (32)\\
%Talk: Large-scale star formation and the bow-shock trigger scenario \\
%Poster: Stellar content of supergiant shells in the LMC\\
%Talk \& Poster --> The young large-scale features in the Large Magellanic
%                   Cloud \\
Sternwarte der\\ Universit{\"a}t Bonn\\
{\tt jbraun@astro.uni-bonn.de} \titsep

{\bf Michael Braun} \plns (63)\\
%Talk: A detailed view of the Magellanic Clouds in the FIR and the possible
%      interaction of HVCs with the disk of the LMC\\
ESA-VILSPA Villafranca del Castillo\\
Satellite Tracking Station\\
{\tt mbraun@iso.vilspa.esa.es} \titsep

{\bf Torbj{\o}rn Bremnes} \plns (29)\\
%Poster: Structure and stellar content of dwarf galaxies: Surface photometry
%        of dwarf galaxies in the M~81 and M~101 groups\\
Astronomisches Institut der\\ Universit{\"a}t Basel\\
{\tt bremnes@astro.unibas.ch} \titsep

{\bf Elias Brinks} \plns ({\bf 39}) \\
% Review: The Violent Interstellar Medium in Dwarf Galaxies: Atomic Gas \\
Departamento de Astronomia, \\
Universidad de Guanajuato \\
{\tt ebrinks@andromeda.cimat.mx} \titsep

% ****************************************************************************
\end{minipage} \qquad
\begin{minipage}[t]{7.6cm}

{\bf Christian Br{\"u}ns} \plns (50)\\
%Poster: \HI\ Velocity Bridges in the Magellanic Stream\\
Radioastronomisches Institut der\\ Universit{\"a}t Bonn\\
{\tt cbruens@astro.uni-bonn.de} \titsep

{\bf Luz Marina Cair{\'o}s Barreto} \plns (57)\\
%Poster: The Low Surface Brightness  Component in dwarf galaxies:
%        evidence of an older population.\\
Instituto de Astrof\'{\i}sica\\ de Canarias (IAC)\\
{\tt lcairos@ll.iac.es} \titsep

{\bf Rolf Chini} \plns (62)\\
Astronomisches Institut der\\
Ruhr-Universit{\"a}t Bochum\\
{\tt chini@astro.ruhr-uni-bochum.de} \titsep

{\bf You-Hua Chu} \plns ({\bf -}) \\
% Review: Hot Gas in the LMC \\
Astronomy Department,\\ University of Illinois \\
{\tt chu@astro.uiuc.edu} \titsep

{\bf Krzysztof Tadeusz Chy\.{z}y} \plns (23)\\
%Poster: Recent radio and polarisation observations of dwarf irregulars\\
Astronomical Observatory,\\ Jagiellonian University\\
{\tt uochyzy@cyf-kr.edu.pl} \titsep

{\bf Maria Rosa Cioni} \plns (20)\\
%Poster: DENIS observations of the Magellanic Clouds\\
Sterrewacht Leiden \\
{\tt mrcioni@strw.leidenuniv.nl} \\
\titsep

{\bf Oliver-Mark Cordes} \plns (60)\\
Sternwarte der\\ Universit{\"a}t Bonn\\
{\tt ocordes@astro.uni-bonn.de} \titsep

{\bf Klaas S. de Boer} \plns (43)\\
%Talk: Bow shock induced star formation in the Large Magellanic Cloud\\
Sternwarte  der\\ Universit{\"a}t Bonn\\
{\tt deboer@astro.uni-bonn.de} \titsep

{\bf Ralf-J{\"u}rgen Dettmar} \plns (71)\\
Astronomisches Institut der\\
Ruhr-Universit{\"a}t Bochum\\
{\tt dettmar@astro.ruhr-uni-bochum.de} \titsep

{\bf Andrea Dieball} \plns (54)\\
%Talk: Binary star clusters in the LMC\\
%Poster: Studies of binary clusters: SL~538 / NGC~2006\\
Sternwarte der\\ Universit{\"a}t Bonn\\
{\tt adieball@astro.uni-bonn.de} \titsep

% ****************************************************************************
\end{minipage}
\newpage

\noindent
\begin{minipage}[t]{7.6cm}

{\bf Pierre-Alain Duc} \plns (44)\\
%Talk: Properties of tidal dwarf galaxies:
%      implications for dwarfs in clusters \\
European Southern Observatory\\ (ESO), Garching\\
{\tt pduc@eso.org} \titsep

{\bf Christian D{\"u}sterberg} \plns (70)\\
%Poster: Temperatures of \HI\ clouds in the
%        30~Dor region AND Shells and Holes in the 30~Dor region\\
Radioastronomisches Institut der\\ Universit{\"a}t Bonn\\
{\tt cduester@astro.uni-bonn.de} \titsep

{\bf So\v{n}a Ehlerov\'a} \plns (1)\\
%Poster :  \HI\ shells in the Milky Way\\
Astronomical Institute,\\ Academy of Sciences of the Czech Republic\\
{\tt sona@ig.cas.cz} \titsep

{\bf Rebecca A.W. Elson} \plns (38)\\
%Poster: 1. An HST Study of Rich Star Clusters in the LMC \\
%        2. The Local Group at High Redshift \\
Institute of Astronomy,\\ Cambridge\\
{\tt elson@ast.cam.ac.uk} \titsep

{\bf Miroslav D. Filipovi{\'c}} \plns (66)\\
%Talk: ATCA Continuum Observations of SNRs in the Magellanic Clouds\\
University of Western Sydney Nepean  / \\
MPE, Garching\\
% Max-Planck-Institut f{\"u}r extraterrestrische Physik (MPE) \\
{\tt fica@st.nepean.uws.edu.au} \titsep

{\bf Thomas Fritz} \plns (49)\\
%Poster: 1. Dwarf galaxies in the Centaurus Group -
%           Interaction with the Environment\\
%        2. Molecular Gas in the BCDGs UM~465 and Haro~2\\
Radioastronomisches Institut der\\ Universit{\"a}t Bonn\\
{\tt tfritz@astro.uni-bonn.de} \titsep

{\bf Uta Fritze - v. Alvensleben} \plns (45)\\
%Talk: Tidal Tail Dwarf Galaxies: Their Present and Future State of Evolution\\
Universit{\"a}ts-Sternwarte\\ G{\"o}ttingen\\
{\tt ufritze@uni-sw.gwdg.de} \titsep

{\bf John S. Gallagher} \plns ({\bf 24}) \\
% Review: Astrophysics in the Dwarf Galaxy Zoo \\
Department of Astronomy,\\ University of Wisconsin at Madison \\
{\tt jsg@tiger.astro.wisc.edu} \titsep

{\bf Carme Gallart} \plns (42)\\
%Talk: The star formation history of Leo~I: A dwarf galaxy with little
%      old population\\
%Poster: A 'bump' above the red-clump in the color-magnitude diagram of
%        nearby galaxies: the AGB-bump\\
Carnegie Observatories \\
{\tt carme@ociw.edu} \titsep

{\bf Eva K. Grebel} \plns (68)\\
%Talk: The Recent Star Formation History of the LMC\\
%Poster:   1. Be stars and IMF of young clusters \\
%          2. First Results from the Magellanic Cloud Photometric Survey
UCO/Lick Observatory,\\ University of California at Santa Cruz \\
%UCSC\\
{\tt grebel@ucolick.org} \titsep

{\bf Martin Haas} \plns (4)\\
%Talk: Cold dust in the peculiar dwarf elliptical galaxy NGC205\\
Max-Planck-Institut f{\"u}r Astronomie\\ (MPIA), Heidelberg\\
{\tt haas@mpia-hd.mpg.de} \titsep

% ****************************************************************************
\end{minipage} \qquad
\begin{minipage}[t]{7.6cm}

{\bf Harm Jan Habing} \plns (48)\\
Sterrewacht Leiden,\\ Huygens Laboratorium\\
{\tt habing@strw.leidenuniv.nl} \titsep

{\bf Andreas Heithausen}\\
Radioastronomisches Institut der\\ Universit{\"a}t Bonn\\
{\tt heith@astro.uni-bonn.de} \titsep

{\bf Fabian Heitsch} \plns (72)\\
Sternwarte der\\ Universit{\"a}t Bonn\\
{\tt fheitsch@astro.uni-bonn.de} \titsep

{\bf Ana Maria Hidalgo-G{\'a}mez} \plns (58)\\
%Talk: The chemical abundances of some nearby dwarf irregular galaxies\\
%Poster: Does the irregular galaxy IC 2574 obey MOND?\\
Astronomiska Observatoriet,\\ Uppsala\\
{\tt anamaria@astro.uu.se} \titsep

{\bf Michael Hilker} \plns (31)\\
%Poster: The center of the Fornax cluster: dwarf galaxies, cD halo,
%        and globular clusters\\
% formerly:  Sternwarte der\\ Universit{\"a}t Bonn\\
% formerly:  {\tt mhilker@astro.uni-bonn.de} \titsep
Departamento de Astronom\'\i a y Astrof\'\i sica, \\
P.~Universidad Cat\'olica, Santiago, Chile\\
{\tt mhilker@astro.puc.cl} \titsep

{\bf Dirk Hoffmann} \plns (15)\\
%Poster: Photometric identification of Be stars in different clusters of the
%        Magellanic Clouds \\
Astronomisches Institut der\\
Ruhr-Universit{\"a}t Bochum\\
{\tt dhoffman@astro.ruhr-uni-bochum.de} \titsep

{\bf Ulrich Hopp} \plns (14)\\
%Talk: The stellar census of the nearby BCDG VII Zw 403 with HST\\
Universit{\"a}ts-Sternwarte\\ M{\"u}nchen \\
{\tt hopp@usm.uni-muenchen.de} \titsep

{\bf Walter Huchtmeier} \plns (17)\\
%Talk: \HI\ properties of new nearby dwarf galaxies
%      from the Karachentsev catalog\\
%Poster: Neutral Hydrogen in a sample of selected BCDG's \\
Max-Planck-Institut f{\"u}r Radioastronomie\\ (MPIfR), Bonn\\
{\tt huchtmeier@mpifr-bonn.mpg.de} \titsep

{\bf Deidre A. Hunter} \plns ({\bf 46}) \\
% Review: Star-forming regions and ionized gas in irregular galaxies \\
Lowell Observatory, Flagstaff \\
{\tt dah@lowell.Lowell.Edu} \titsep

{\bf Susanne H{\"u}ttemeister}\\
Radioastronomisches Institut der\\ Universit{\"a}t Bonn\\
{\tt huette@astro.uni-bonn.de} \titsep

{\bf Norbert Junkes} \plns (61)\\
%Talk: X-ray Observations of Starburst Dwarf Galaxies: The Cases of
% NGC~1705 and He2-10\\
% formerly:  Astrophysikalisches Institut Potsdam\\ (AIP)\\
% formerly:  {\tt njunkes@aip.de} \titsep
Max-Planck-Institut f\"ur Radioastronomie \\ (MPIfR), Bonn\\
{\tt njunkes@mpifr-bonn.mpg.de} \titsep

% ****************************************************************************
\end{minipage}
\newpage

\noindent
\begin{minipage}[t]{7.6cm}

{\bf Marcus J{\"u}tte} \plns (34)\\
%Poster: NIR imaging of dwarf galaxies\\
Astronomisches Institut der\\
Ruhr-Universit{\"a}t Bochum\\
{\tt juette@astro.ruhr-uni-bochum.de} \titsep

{\bf Sungeun Kim} \plns (41)\\
%Talk: An \HI\ aperture synthesis mosaic survey of the Large Magellanic Cloud\\
Mount Stromlo and Siding Spring\\ Observatories \\
{\tt sek@mso.anu.edu.au} \titsep

{\bf Uli Klein} \plns (36)\\
Radioastronomisches Institut der\\ Universit{\"a}t Bonn\\
{\tt uklein@astro.uni-bonn.de} \titsep

{\bf Hans-Rainer Kl\"ockner}\\
% Poster: Comparison of H<SMALL>&#160;I</SMALL> and magnetic field structures
%         in our Galaxy<BR>
Radioastronomisches Institut der\\ Universit{\"a}t Bonn\\
{\tt hrkloeck@astro.uni-bonn.de} \titsep

{\bf Joachim K\"oppen} \plns ({\bf -})\\
% Review: Star formation law in chemodynamical models and disk galaxies\\
Equipe 'Evolution Galacticque', \\ Observatoire de Strasbourg\\
{\tt koppen@astro.u-strasbg.fr} \titsep

{\bf Sven Kohle} \plns (53)\\
%Poster: A complete CO survey of NGC~4449\\
Radioastronomisches Institut der\\ Universit{\"a}t Bonn\\
{\tt skohle@astro.uni-bonn.de} \titsep

{\bf Pavel Kroupa} \plns (33)\\
%Talk: dSph satellite galaxies without dark matter: a study of parameter space\\
Institut f{\"u}r Theoretische Astrophysik\\ der Universit{\"a}t Heidelberg\\
{\tt pavel@ita.uni-heidelberg.de} \titsep

{\bf Harald Lesch} \plns (19)\\
%Talk: Dwarf galaxies and intergalactic magnetic fields\\
Universit{\"a}ts-Sternwarte\\ M{\"u}nchen\\
{\tt lesch@usm.uni-muenchen.de} \titsep

{\bf Rainer L{\"u}tticke} \plns (51)\\
%Poster: Influence of satellite companions on the evolution of box/peanut
%        bulges\\
Astronomisches Institut der\\
Ruhr-Universit{\"a}t Bochum\\
{\tt luett@astro.ruhr-uni-bochum.de} \titsep

{\bf Mordecai-Mark Mac Low} \plns (3)\\
%Talk: Starbursts in Dwarf Galaxies:  Blown Out or Blown Away?\\
%Poster: ROSAT observations of the giant \HII\ complex N~11 in the LMC\\
Max-Planck-Institut f{\"u}r Astronomie\\ (MPIA), Heidelberg\\
{\tt mordecai@mpia-hd.mpg.de} \titsep

{\bf David Mart\'\i nez-Delgado} \plns (55)\\
%Poster: The star formation history of the Local Group dwarf galaxy Phoenix\\
Instituto de Astrof\'{\i}sica\\ de Canarias (IAC)\\
{\tt ddelgado@ll.iac.es} \titsep

% ****************************************************************************
\end{minipage} \qquad
\begin{minipage}[t]{7.6cm}

{\bf Monika Marx-Zimmer}\\
%Poster: 1. Properties of the Cool Atomic Gas in the LMC -
%           a third \HI\ Absorption Survey\\
%        2. CO Emission in Direction to Cool \HI\ Clouds in the LMC\\
%        3. {C$\,${\sc ii}} Investigation of Cool Clouds in the LMC
Radioastronomisches Institut der\\ Universit{\"a}t Bonn\\
{\tt mmarx@astro.uni-bonn.de} \titsep

{\bf Mario L. Mateo} \plns ({\bf 8}) \\
% Review: Strange dark matters in dwarf galaxies \\
Department of Astronomy,\\ University of Michigan \\
{\tt mateo@astro.lsa.umich.edu} \titsep

{\bf Ulrich Mebold}\\
Radioastronomisches Institut  der\\ Universit{\"a}t Bonn\\
{\tt mebold@astro.uni-bonn.de} \titsep

{\bf David Israel M{\'e}ndez Alcaraz} \plns (56)\\
%Talk: Star formation in dwarf wolf-rayet galaxies \\
Instituto de Astrof\'{\i}sica\\ de Canarias (IAC)\\
{\tt dmendez@ll.iac.es} \titsep

{\bf Andrea Moneti} \plns (30)\\
ESA-VILSPA Villafranca del Castillo\\
Satellite Tracking Station\\
{\tt amoneti@iso.vilspa.esa.es} \titsep

{\bf Kai Noeske} \plns (28)\\
%Poster: Multi-Color Surface Photometry of Blue Compact Dwarf Galaxies\\
Universit{\"a}ts-Sternwarte\\ G{\"o}ttingen\\
{\tt knoeske@uni-sw.gwdg.de} \titsep

{\bf Marion Siang-Li Oey} \plns (74)\\
%Talk: Interpreting the \HII\ Region Luminosity Function \\
Institute of Astronomy,\\ Cambridge\\
{\tt oey@ast.cam.ac.uk} \titsep

{\bf Jan Palou\v{s}} \plns ({\bf 10}) \\
% Review: Holes and shells in galaxies:
% observeation versus theoretical concepts \\
Astronomical Institute, Academy of \\ Sciences of the Czech Republic \\
{\tt PALOUS@ig.cas.cz} \titsep

{\bf Nino Panagia} \plns (7)\\
%Talk: HST Study of the Stellar Population within 30~pc of SN~1987A\\
% M. Romaniello (STScI/SNS-Pisa), N. Panagia (STScI/ESA), S. Scuderi
% (Obs. Catania), and the SINS Collaboration.\\
Space Telescope Science Institute\\ (STScI), Baltimore\\
{\tt panagia@stsci.edu} \titsep

{\bf Daniele Pierini} \plns (18)\\
%Poster: The K'-band TF relation of Virgo late-type galaxies.\\
Max-Planck-Institut f{\"u}r Kernphysik,\\ Heidelberg\\
{\tt pierini@peter.mpi-hd.mpg.de} \titsep

{\bf Sean Points} \plns (52)\\
%Poster: The Kinematic and Physical Structure of the Supergiant Shell LMC~2\\
Astronomy Department,\\ University of Illinois\\
{\tt points@astro.uiuc.edu} \titsep

% ****************************************************************************
\end{minipage}
\newpage

\noindent
\begin{minipage}[t]{7.6cm}

{\bf Gotthard Richter}\\
Astrophysikalisches Institut Potsdam\\ (AIP)\\
{\tt gmrichter@aip.de} \titsep

{\bf Philipp Richter} \\
%Talk and Poster: UV spectroscopy for two lines of sight to the LMC\\
Sternwarte der\\ Universit{\"a}t Bonn\\
{\tt prichter@astro.uni-bonn.de} \titsep

{\bf Tom Richtler} \plns (16)\\
%Poster: Primordial mass segragation in young Magellanic Cloud clusters\\
Sternwarte der\\ Universit{\"a}t Bonn\\
{\tt richtler@astro.uni-bonn.de} \titsep

{\bf Andreas Rieschick} \plns (64)\\
%Talk: Chemodynamical Evolution of Dwarf Galaxies: New Simulations\\
Institut f{\"u}r Theoretische Physik und\\ Astrophysik, Kiel\\
{\tt rieschick@astrophysik.uni-kiel.de} \titsep

{\bf Martino Romaniello} \plns (69)\\
%Talk: HST Study of the Stellar Populations around SN1987A\\
% Space Telescope Science Institute
STScI, Baltimore, USA and\\ Scuola Normale Superiore, Pisa, Italy\\
{\tt martino@cibs.sns.it} \titsep

{\bf Pierre Royer} \plns (47)\\
%Poster: A photometric method to study the Wolf-Rayet content of compact
%        \HII\ regions in nearby galaxies\\
Institut d'astrophysique de Li{\`e}ge \\
{\tt proyer@ulg.ac.be} \titsep

{\bf Theodor Schmidt-Kaler}\\
%Poster: 1. CCD Photometry of Bright Multiple Stars (Superginats) and the
%           Association LH~10 in the LMC\\
%        2. Integrated Colours (<I>U-B</I>), (<I>B-V</I>), (<I>V-R</I>) and
%           Magnitude <I>V</I> of the LMC from Surface Photometry\\
%        3. Photoelectric <I>UBV</I> Photometry of LMC Member Stars
%           - a New Database
Astronomisches Institut der\\
Ruhr-Universit{\"a}t Bochum\\
\titsep

{\bf Uwe Schwarzkopf} \plns (6)\\
%Poster: Interacting and merging processes between spirals and dwarf galaxies\\
Astronomisches Institut der\\
Ruhr-Universit{\"a}t Bochum\\
{\tt schwarz@astro.ruhr-uni-bochum.de} \titsep

{\bf Wilhelm Seggewiss} \plns (73)\\
Sternwarte der Universit{\"a}t Bonn / \\ Observatorium Hoher List \\
{\tt seggewis@astro.uni-bonn.de} \titsep

{\bf Evan D. Skillman} \plns ({\bf 27}) \\
% Review: Star formation histories of nearby dwarf galaxies \\
Astronomy Department,\\ University of Minnesota \\
{\tt skillman@astro.spa.umn.edu} \titsep

% ****************************************************************************
\end{minipage} \qquad
\begin{minipage}[t]{7.6cm}

{\bf Chris Taylor} \plns (21)\\
%Talk: 1. The Role of Galaxy Interactions in \HII\ Galaxies\\
%      2. CO emission in low luminosity star forming galaxies
Astronomisches Institut der\\
Ruhr-Universit{\"a}t Bochum\\
{\tt taylorc@astro.ruhr-uni-bochum.de} \titsep

{\bf Christian Theis} \plns (65)\\
%Talk: The outer halo of NGC~4449\\
Institut f{\"u}r Theoretische Physik und\\ Astrophysik, Kiel\\
{\tt theis@astrophysik.uni-kiel.de} \titsep

{\bf Trinh X. Thuan} \plns ({\bf 26}) \\
% Review: Young dwarf galaxies \\
Department of Astronomy,\\ University of Virginia \\
{\tt txt@starburst.astro.virginia.edu} \titsep

{\bf Eline Tolstoy} \plns (25)\\
%Talk: Leo~A: a truly young galaxy?\\
European Southern Observatory\\ (ESO), Garching\\
{\tt etolstoy@eso.org} \titsep

{\bf Monica Tosi} \plns (22)\\
%Talk: Star formation history in the post-starburst galaxy NGC1569\\
Osservatorio Astronomico\\ di Bologna \\
{\tt tosi@astbo3.bo.astro.it} \titsep

{\bf Hugo van Woerden} \plns (59)\\
Kapteyn Institute, Groningen\\
{\tt hugo@astro.rug.nl} \titsep

{\bf Fabian Walter} \plns (35)\\
%Talk: The Violent Interstellar Medium of IC 2574\\
%Poster: 1. Holes and Shells in IC~2574\\
%        2. \HI\ study of the dwarf galaxy DDO~47\\
Radioastronomisches Institut der\\ Universit{\"a}t Bonn\\
{\tt walter@astro.uni-bonn.de} \titsep

{\bf Kerstin Weis} \plns (2)\\
% Poster: Very Massive Stars in the Magellanic Clouds \HII\ regions\\
Institut f{\"u}r Theoretische Astrophysik,\\ Heidelberg\\
{\tt kweis@ita.uni-heidelberg.de} \titsep

{\bf Axel Weiss} \plns (37)\\
Radioastronomisches Institut der\\ Universit{\"a}t Bonn\\
{\tt aweiss@astro.uni-bonn.de} \titsep

{\bf Bernhard Wierig} \plns (9)\\
% Poster: Spatial distribution of halo high velocity clouds towards the LMC<BR>
Sternwarte der\\ Universit{\"a}t Bonn\\
{\tt bwierig@astro.uni-bonn.de} \titsep

% ****************************************************************************
\end{minipage}
\newpage

\noindent
\begin{minipage}[t]{7.6cm}

{\bf Hans Zinnecker}\\
Astrophysikalisches Institut Potsdam\\ (AIP)\\
{\tt hzinnecker@aip.de} \titsep

% ****************************************************************************
\end{minipage} \qquad
\begin{minipage}[t]{7.6cm}

% If there are some people missing in this list ...

\end{minipage}
\titsep

\noindent
\rule{4.67cm}{0.02cm} \vspace*{0.15cm}

\hspace*{-0.60cm} {\footnotesize \begin{minipage}{16.0cm}
% For the desired text indent:
\setlength{\parindent}{0.5cm}
The list above shows the {\bf names of the participants},
the numbers belonging to the photograph on p. 311
(in brackets, ciphers for invited speakers additionally set in {\bf boldface}),
the names of the institutes and the {\tt E-Mail addresses}.
The full list of participants including all links is available at the WWW URL:
{\tt http://www.astro.uni-bonn.de/$\,\,\tilde{}\,\,$webgk/ws98/part\_a\_tbl.html}.
\end{minipage}}
\vspace*{6cm}

\centerline{
% \hspace*{-0.25cm}
\epsfxsize=6.16cm
% 2.0531cm would use 600 dpi
\epsffile{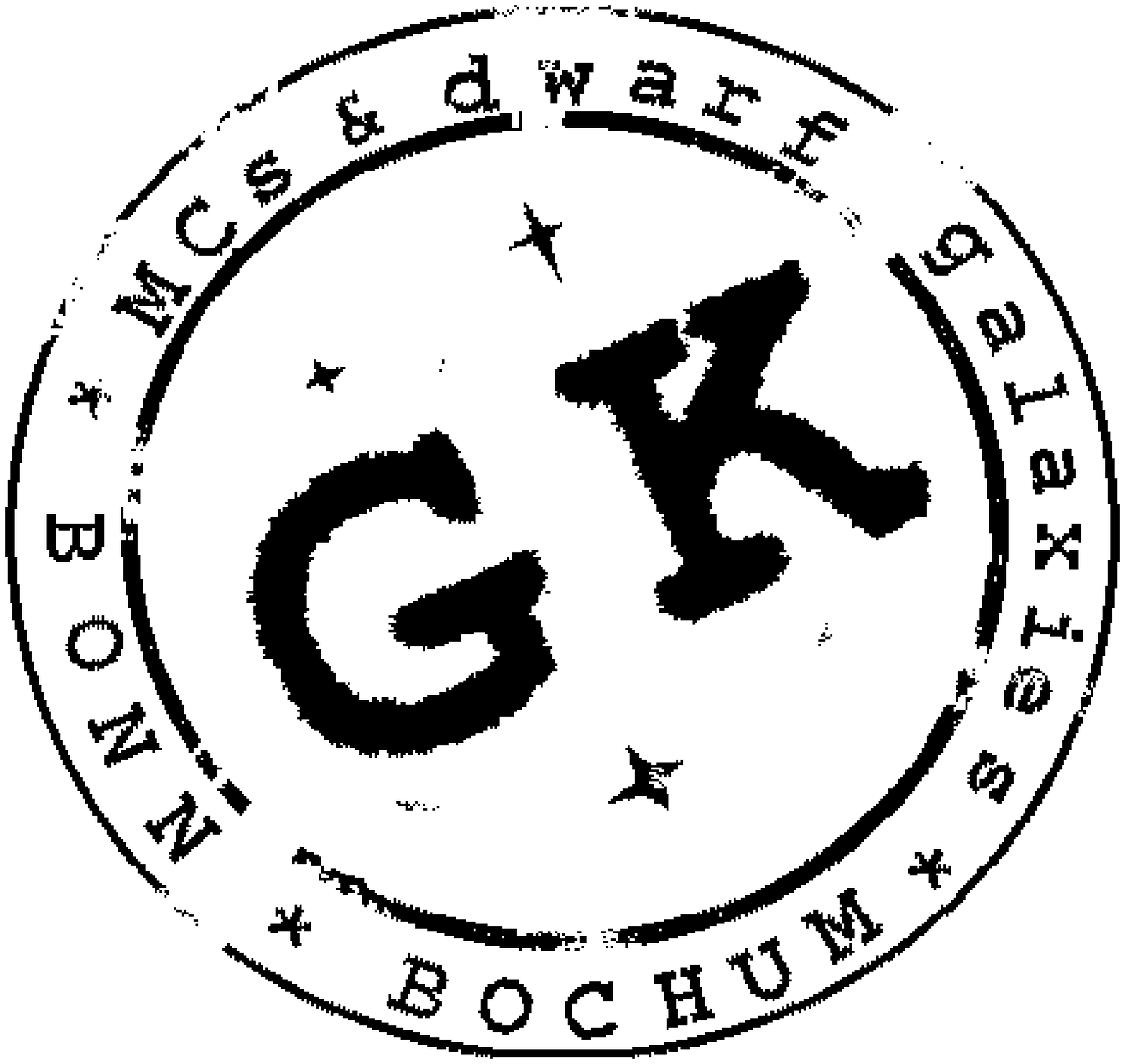}
}

\end{document}